\pdfoutput=1
\documentclass[12pt]{iopart}
\usepackage[latin1]{inputenc}
\usepackage{setspace}
\usepackage{graphicx}
\usepackage{iopams}
\usepackage{amstext}
\usepackage{epstopdf}

\begin{document}

\title{Assessing the accuracy of projected entangled-pair states on infinite lattices}

\author{B. Bauer$^1$, G. Vidal$^2$, M. Troyer$^1$}
\address{$^1$Theoretische Physik, ETH Zurich, 8093 Zurich, Switzerland}
\address{$^2$School of Physical Sciences, The University of Queensland, QLD 4072, Australia}

\date{\today}

\begin{abstract}
Generalizations of the density-matrix renormalization group method have long been sought after. In this paper, we assess the accuracy of projected entangled-pair states on infinite lattices by comparing with Quantum Monte Carlo results for several non-frustrated spin-1/2 systems. Furthermore, we apply the method to a frustrated quantum system.
\end{abstract}


\maketitle

\section{Introduction}
\subsection{Tensor network states}
While the density matrix renormalization group (DMRG) \cite{white1992} has proven to be a very useful numerical tool for the study of strongly correlated quantum many-body systems in one dimension, applications to higher dimensions have remained very limited so far because an exponential growth of the complexity with system size is recovered in general. Monte Carlo simulations, on the other hand, are not limited by the dimensionality of the problem, but perform well only for non-frustrated bosonic systems. Extensions of the DMRG method to two dimensions are therefore highly desired.

In recent years, insights from quantum information theory have led to an understanding of the success of DMRG in terms of entanglement of the ground state: the renormalization procedure of choosing the dominant eigenvectors of the reduced density matrix picks out states with a large contribution to the entanglement entropy. It can therefore be regarded as a low-entanglement ansatz for the ground state; the entanglement that can be encoded is bounded by $S \leq \log M$, where $M$ denotes the number of states kept in a renormalization step. Since for non-critical systems the amount of entanglement saturates as a function of system size \cite{vidal2003}, a finite $M$ will be able to capture the physics of such systems even in the thermodynamic limit. For weakly entangled systems, the ansatz will be a very accurate description even for very small $M$.

Since the block entanglement does not saturate in the thermodynamic limit for two-dimensional systems, the matrix size needs to grow exponentially with the system size in an approach based on DMRG to obtain a fixed accuracy \cite{liang1994}. However, one can hope that the entanglement grows less than the volume of the system; indeed, it has been shown that for many classes of interesting strongly correlated systems an \emph{area law} holds \cite{eisert2008}, i.e. that the entanglement between a given block within a system and the rest of the system will scale with the surface between the two. Therefore, algorithms that obey an area law by construction appear as a viable approach to higher-dimensional systems. One example for such an algorithm will be the topic of this paper.

DMRG can be understood as a variational method operating on \emph{matrix product states} \cite{ostlund1995}. For states in a Hilbert space spanned by an orthonormal basis $|\mathbf{n}\rangle = \prod_{i=1\ldots L} (a_i^{\dagger})^{n_i} | \Omega \rangle$, where the $a_i^{\dagger}$ are appropriate creation operators, $\mathbf{n} = n_1 \ldots n_L$ is an occupation number vector and $| \Omega \rangle$ is the vacuum, the coefficients $c(\mathbf{n})$ of the wave function $| \Psi \rangle = \sum_{\mathbf{n}} c(\mathbf{n}) |\mathbf{n}\rangle$ are expanded as product of a set of matrices of size $M \times M$, where one matrix is associated with each lattice site:
\begin{equation}
c(\mathbf{n}) = A^1_{\alpha_1}[n_1] A^2_{\alpha_1 \alpha_2}[n_2] \ldots A^N_{\alpha_N}[n_N]
\end{equation}
In the limit of infinite matrix size, the set of matrix-product states is complete. It has been shown that MPS offer an efficient description of local and non-local properties of ground states of gapped and even critical systems \cite{verstraete2006-2}.

\begin{figure}[t]
 \centering
   \includegraphics[width=8cm]{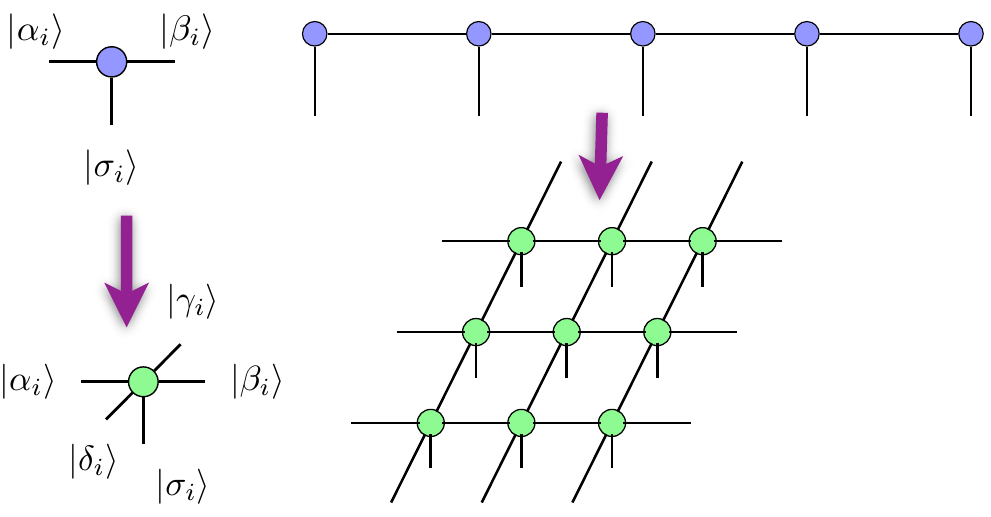}
   \caption{(color online) From \emph{matrix-product states} to \emph{projected entangled-pair states}. In both cases, a linear operator mapping from auxiliary spaces on the bonds is associated with each site. \label{fig:peps}}
\end{figure}

For higher-dimensional systems, several ansatz states and algorithms have been proposed that capture an area law by construction. We will focus on an ansatz that was first suggested by G. Sierra and M. A. Martin-Delgado under the name of vertex matrix product ansatz \cite{sierra1998} and subsequently applied to evaluate the partition function of a 3D classical model by Nishino and Okunishi \cite{nishino1998,nishino2000}. A first application to infinite, homogeneous quantum systems is found in \cite{nishio2004}. A formalism for inhomogeneous, finite systems was proposed by Verstraete and Cirac \cite{verstraete2004} under the name of \emph{projected entangled-pair states}. In \cite{jordan2008}, infinite systems were reconsidered from the point of view of extending Verstraete and Cirac's formalism. For this paper, we choose to refer to the ansatz as PEP states.

Since it was first proposed, many studies of classical and quantum systems using this family of states have appeared \cite{nishino2001,maeshima2001,gendiar2003,gendiar2005,isacsson2006,murg2007,jordan2009,orus2009,orus2009-1,murg2009}. Additionally, other algorithms operating on these states have been proposed, including \emph{tensor entanglement renormalization group} (TERG) \cite{gu2008,jiang2008,chen2009}. Related proposals include the \emph{multiscale entanglement renormalization ansatz} (MERA) \cite{vidal2007-1}, \emph{string-bond states} \cite{schuch2008} and \emph{entangled-plaquette states} \cite{mezzacapo2009}.

\subsection{Projected entangled-pair states}
In the following, we will describe PEP states, which can be understood easily from a diagrammatic representation (Fig. \ref{fig:peps}). For spins distributed on an arbitrary graph, a tensor is associated with each vertex; the tensor has one index for each edge emanating from it plus one physical index (a specific choice of basis is implied). Specializing to the square lattice in two dimensions, we have rank 5 tensors. In the following, we will choose the dimensions of all indices between tensors equal to $M$ and denote the dimension of the physical Hilbert space by $d$. The bond dimension $M$ provides a systematic refinement parameter: while for ground state properties, an $M=1$ PEPS corresponds to a mean-field state, the class of states is complete in the limit of $M \rightarrow \infty$. The values of $M$ attainable in practical calculations are very limited due to a strong, but polynomial increase of the computation time with $M$. On the other hand, the fact that for increasing dimension of the lattice, typical ground states of local Hamiltonians are closer to mean-field states suggests that the dimension of the bonds in a PEPS can be chosen smaller than in an MPS. Nevertheless, this is the most serious limitation to the accuracy of the algorithm. One of the main purposes of this paper will be to assess the accuracy obtained with PEPS for the values of $M$ attained in practical calculations.

For a wide variety of systems, the PEPS may provide a very efficient representation of the exponential number of coefficients with a small number of parameters. However, the aforementioned results on the accuracy of MPS for gapped and even critical systems cannot be extended to PEPS. Additionally, it can be shown that the exact calculation of an expectation value based on such a network, i.e. the trace over all indices, is an exponentially hard problem and therefore intractable directly for larger system sizes \cite{schuch2007-1}. To overcome this limitation, approximate contraction schemes have been devised. We will discuss the \emph{infinite projected entangled-pair states algorithm} (iPEPS), which operates directly in the thermodynamic limit.

The treatment of the infinite lattice depends on the observation that the infinite tensor network can be regarded as transfer operator acting on some boundary state and all correlators can be expressed in terms of the dominant eigenvectors of this transfer operator. This represents a renormalization scheme that replaces expectation values of local operators with expectation values of effective operators that take into account long-range correlations, but can be evaluated locally.

It should be emphasized that we do not handle boundary conditions explicitly. In some cases, the boundary state that we use to locally approximate expectation values becomes degenerate, in which case a choice of boundary condition is implied by the choice of boundary state from the manifold of degenerate states. For systems with a broken local symmetry, this degeneracy is related to the degeneracy of states with different symmetry breaking direction. We however predetermine this direction by the choice of the initial state, which is not symmetric and therefore favours a symmetry sector. In the case of topological order, this situation is much more intricate. For the purpose of this paper, we therefore restrict ourselves to systems where only local symmetries are relevant. Systems with topological order have been treated elsewhere, e.g. \cite{gu2009,buershaper2009,gu2008}.

The outline of the paper is as follows: in section \ref{sect:ipeps}, we will illustrate the details of the infinite projected entangled-pair algorithm and discuss properties of the ansatz in the vicinity of a first-order phase transition. In section \ref{sect:nonfrust}, we apply the algorithm to well-known non-frustrated systems and assess the accuracy of the results in comparison to Quantum Monte Carlo simulations. In section \ref{sct:frust}, we extend the simulations to frustrated systems where Monte Carlo methods fail.

\section{Infinite PEPS algorithm} \label{sect:ipeps}

\begin{figure}[t]
 \centering
   \includegraphics[width=2in]{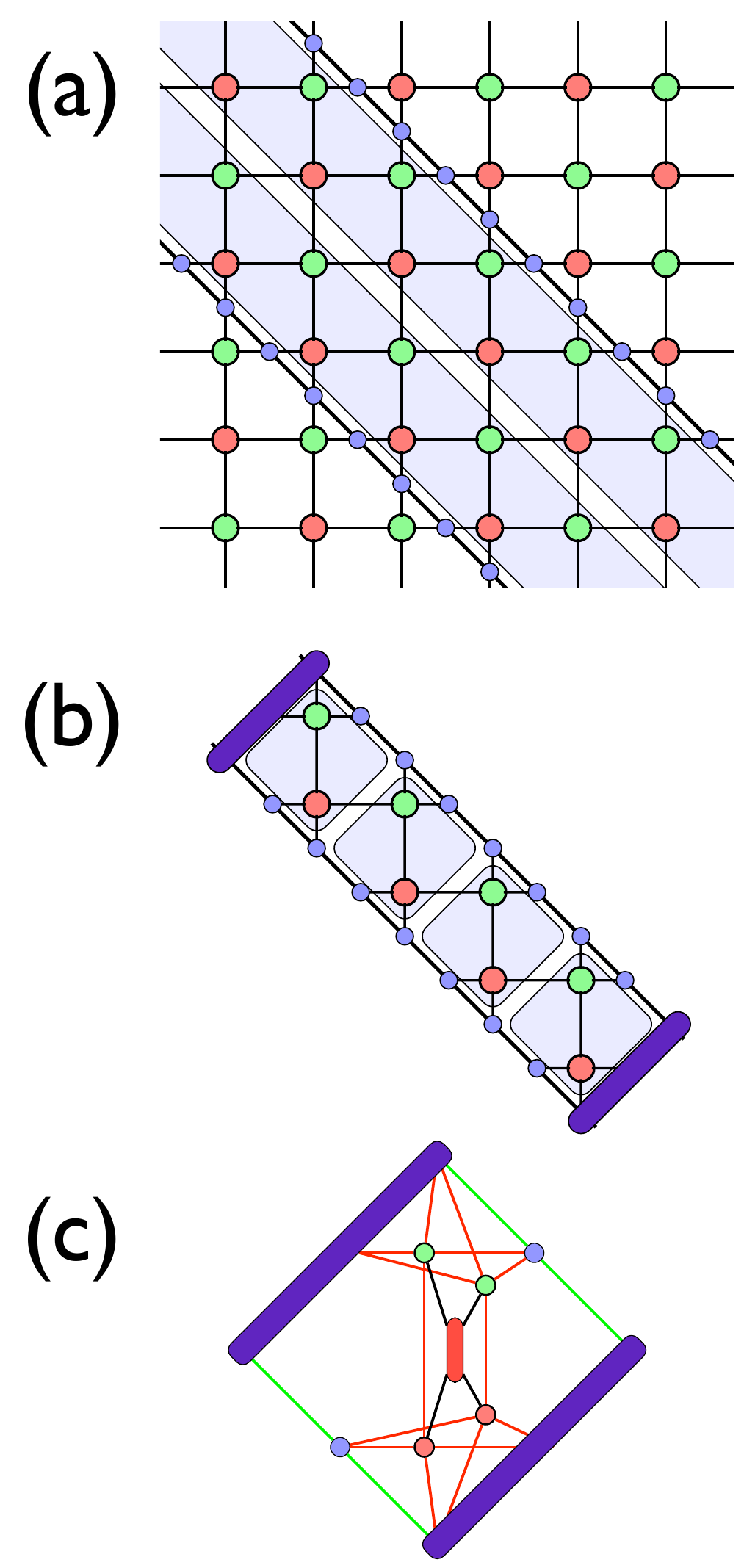}
   \caption{(color online) MPS-based scheme for the reduction of the infinite tensor network to a small network. In this scheme, diagonals or rows/columns of the lattice are regarded as infinite transfer operators and operators are evaluated using the dominant eigenvectors of these operators. We use a diagonal contraction scheme. In (a) and (b), the physical indices have already been traced out. (a) Tensor network with two-tensor unit cell, diagonal transfer operators and boundary iMPS. (b) The two-dimensional lattice is reduced to a quasi-one-dimensional system by approximation with the dominant eigenvectors of the transfer operators. (c) Using the dominant eigenvectors of the one-dimensional strip, a finite network is obtained in which operators can be applied to the physical indices. The trace over this network equals the expectation value of that operator. \label{fig:scheme}}
\end{figure}

\subsection{Approximation scheme}
Several schemes are available for the approximation of the infinite tensor network by a small tensor network in which operators can be evaluated efficiently. This includes a scheme based on boundary infinite matrix-product states as eigenvectors of diagonals or rows/columns of the lattice (cf. Fig. \ref{fig:scheme}) \cite{jordan2008,orus2009,jordan2009} and the corner-transfer matrix scheme originally proposed in the context of tensor-product states by Nishino and Okunishi \cite{nishino1996} and recently revisited by Orus and Vidal \cite{orus2009-1}.

While the direct comparison of the two methods in \cite{orus2009-1} by example of the two-dimensional quantum Ising model in transverse field yields slightly better results for the corner-transfer matrix, this is expected to be due to the fact that a very small bond dimension $M=2,3$ is sufficient to capture the physics of the Ising model. In this paper, we will consider models with more entangled ground states where errors result from insufficient $M$ instead of an inaccurate contraction of the environment. Therefore, we do not expect significant improvement from the use of the corner-transfer matrix. We have checked this for $M=2$ by example of the Heisenberg model (Section \ref{sct:xxxxy}) and the dimerized Heisenberg model (Section \ref{sct:dimerized_hb}) and found the deviations between the boundary MPS method and the CTM to be negligible. In the case of the frustrated model (Section \ref{sct:frust}), we use a slightly different scheme to update the boundary MPS in order to accomodate for the enlarged unit cell. The deviations in this case are slightly larger, but remain within about 2 \% for the order parameter.

\begin{figure}[t]
 \centering
   \includegraphics[width=8cm]{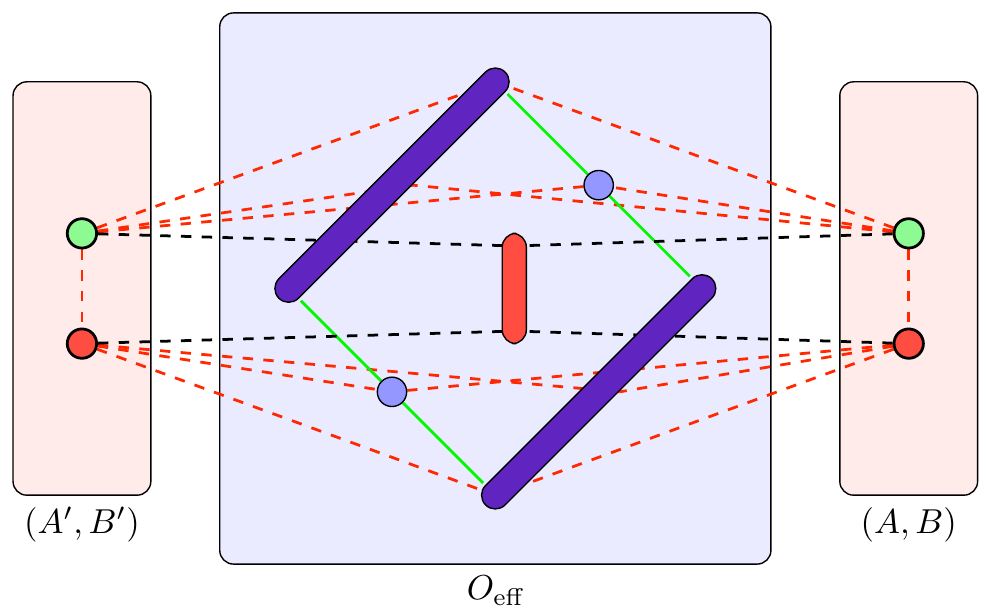}
   \caption{(color online) Graphical representation of eqn. (\ref{eqn:approx}); this network is identical to Fig. \ref{fig:scheme} (c). The effective renormalized operator consists of the two-site operator and the environment representing the infinite lattice. By means of this renormalization, it takes into account long-range correlations in the state. The trace over all bonds yields the expectation value of $O$. \label{fig:eff_operator}}
\end{figure}

Finding the dominant eigenvectors of the infinite transfer operator can be performed in several ways. Most commonly, the iTEBD algorithm for non-unitary evolution \cite{orus2008} is applied. We find that for lattices with a larger unit cell, better numerical stability can be achieved using an algorithm based on the optimization of fidelity without requiring orthogonalization; such an approach was suggested in \cite{verstraete2004-1}. For the boundary of the strip (cf. Fig. \ref{fig:scheme} (b)), it is sufficient to use a single tensor as boundary.

In order to reduce the complexity of operations, transfer matrices are not stored explicitly. Instead, iterative schemes are applied where the matrix-vector product is implemented by exploiting the tensor structure in such a way that only two tensors are contracted at a time, storing the result in a temporary tensor. This operation can be mapped to a matrix multiplication, for which efficient computer libraries exist. The optimal order of contractions has to determined depending on the dimension of each bond.

The local environment provides a means of expressing diagonal or off-diagonal operators $O$ in an effective form acting on PEPS tensors,
\begin{equation} \label{eqn:approx}
\langle \Psi' | O | \Psi \rangle = (A',B')^{\dagger} O_{\text{eff}} (A,B),
\end{equation}
where the $(A,B)$ are PEPS tensors that can be regarded as vectors in $\mathbb{C}^{2M^4d}$. The accuracy to which expectation values of effective operators approximate the exact expectation value is controlled by the bond dimension of the boundary matrix-product states, $M'$. The computational cost of increasing $M'$ is very large, since the iTEBD algorithm requires the singular-value decomposition of a matrix of dimension $M^2 M' \times M^2 M'$, which scales as $M^6 M'^3$; for realistic choices of $M'$ and $M$, where $M' \gtrsim M^2$, this becomes the most time-consuming operation of the algorithm. In the simulations, we choose $M'$ large enough such that the results do not change with increased $M'$; nevertheless, we should mention that errors made at this step enter both into the calculation of the ground state and the evaluation of expectation values.

In order to project the state into the ground state, imaginary time evolution is performed. For a Hamiltonian that can be written as a sum of strictly local terms, $H = \sum_{\langle i,j \rangle} h_{ij}$, the operator $\exp(-\beta H)$ can be expanded straightforwardly by discretizing time $\beta = N \delta \tau$ and applying a Trotter decomposition \cite{trotter1990},
\begin{equation}
U = \exp(-\delta \tau H) = \prod_{\langle i,j \rangle} \exp(-\delta \tau h_{ij} ) + \mathcal{O}(\delta \tau^2).
\end{equation}
In the case of imaginary time evolution, Trotter errors do not accumulate and can largely be eliminated by reducing the time step during the simulation and using higher-order expansions.

Applying the evolution gate directly to the PEPS tensors and transforming back to the previous form increases the bond dimension of the PEPS. Therefore, time evolution can only be approximated by finding the state $|\Psi'\rangle$ that minimizes
\begin{equation} \label{eqn:opti}
D = \parallel U | \Psi \rangle - | \Psi' \rangle \parallel.
\end{equation}
In order to achieve an optimal approximation, the norm $\parallel \cdot \parallel$ has to be chosen in such a way that the correlations in the system are taken into account. To this end, the evolution operator $U$ can be renormalized to $U_{\text{eff}}$ using the scheme discussed above. Equation (\ref{eqn:opti}) can then be written as
\begin{eqnarray}
D &=& \langle \Psi | U^{\dagger} U | \Psi \rangle - \langle \Psi' | U | \Psi \rangle \\
\ &\ & - \langle \Psi | U^{\dagger} | \Psi' \rangle  + \langle \Psi' | \Psi \rangle \\
\ &=& (A,B)^{\dagger} (U^{\dagger} U)_{\text{eff}} (A,B) \\
\ &\ & -(A',B')^{\dagger} U_{\text{eff}} (A,B) \\
\ &\ & -(A,B)^{\dagger} U^{\dagger}_{\text{eff}} (A',B') \\
\ &\ & +(A',B')^{\dagger} \mathbb{I}_{\text{eff}} (A',B')
\end{eqnarray}

Minimizing this in $A'$ and $B'$ leads to a set of coupled linear equations which is solved iteratively by fixing one tensor and optimizing in the other, sweeping back and forth until convergence is reached.

In the following, we have used two different unit cells: a staggered AB unit cell, as shown in Fig. \ref{fig:scheme}, which allows for ferromagnetic and antiferromagnetic order, and an enlarged four-site unit cell, which allows more general types of order (cf. also Fig. \ref{fig:dimerization}). We start each simulation with a large timestep of the order of $\delta \tau \approx 10^{-2}$, which is decreased to $\delta \tau < 10^{-5}$ during the simulation. Convergence is checked not only for the energy, but also for all other local observables that are being evaluated.

As initial state for calculations with $M=2$, we use either a completely random state or a product state, e.g. the N\'{e}el state, with a small amount of disorder added. For calculations with larger $M$, we start from extrapolations of previous results to save computation time.

\subsection{iPEPS at a first-order phase transition} \label{sct:fo}
The iPEPS method is very suitable to first-order phase transitions. In a previous work by Or\'{u}s et al \cite{orus2009}, a phase transition in the anisotropic quantum orbital compass model was studied and the transition point located using adiabatic time evolution. We will take a different approach based on the observation that the dynamics close a first-order phase transition differ fundamentally between simulations on finite lattices and on infinite lattices with translational invariance.

If the system is taken across the phase transition from phase A to phase B, phase A becomes meta-stable. Since phase A cannot locally be continuously deformed into phase B (assuming pure states), the dominant process by which the meta-stable state decays is the condensation of finite clusters of phase B (see, e.g., Binder \cite{binder1987}), thereby breaking translational symmetry. This process is suppressed by enforcing translational symmetry. The imaginary time evolution we simulate therefore has a strongly suppressed probability of tunneling from one phase to the other in the phase coexistence regime where the energy densities of phase A and B are similar.

We can therefore expect to find only homogeneous systems of one phase in the vicinity of the transition. By preparing a state deep within one phase and using this state as initial state for the imaginary time evolution of the Hamiltonian with different parameters, we can find the energy density of each phase for a relatively wide space of parameters. The transition point can then be located to high accuracy as the crossing point of the ground-state energy densities of the two phases; simultaneously, the existence of this stability allows us to clearly distinguish the nature of the phase transition.

\section{Non-frustrated test cases} \label{sect:nonfrust}

We study several non-frustrated spin-$\frac{1}{2}$ models on the square lattice, where results can easily be compared to previous results and Quantum Monte Carlo calculations, which we perform using the ALPS package \cite{ALPS_1,ALPS_2}. As a first test case and complementary to previous results \cite{jordan2009}, we study a first-order phase transition in the XXZ model in external field, which maps to the Bose-Hubbard model in the hard-core limit. We assess the accuracy in the case of gapless, long-range ordered states by example of the XY model and the isotropic Heisenberg model, which we extend to a dimerized model exhibiting a quantum critical point.

\subsection{XXZ model in external field}

\begin{figure}[t] 
   \centering
   \includegraphics[width=8cm]{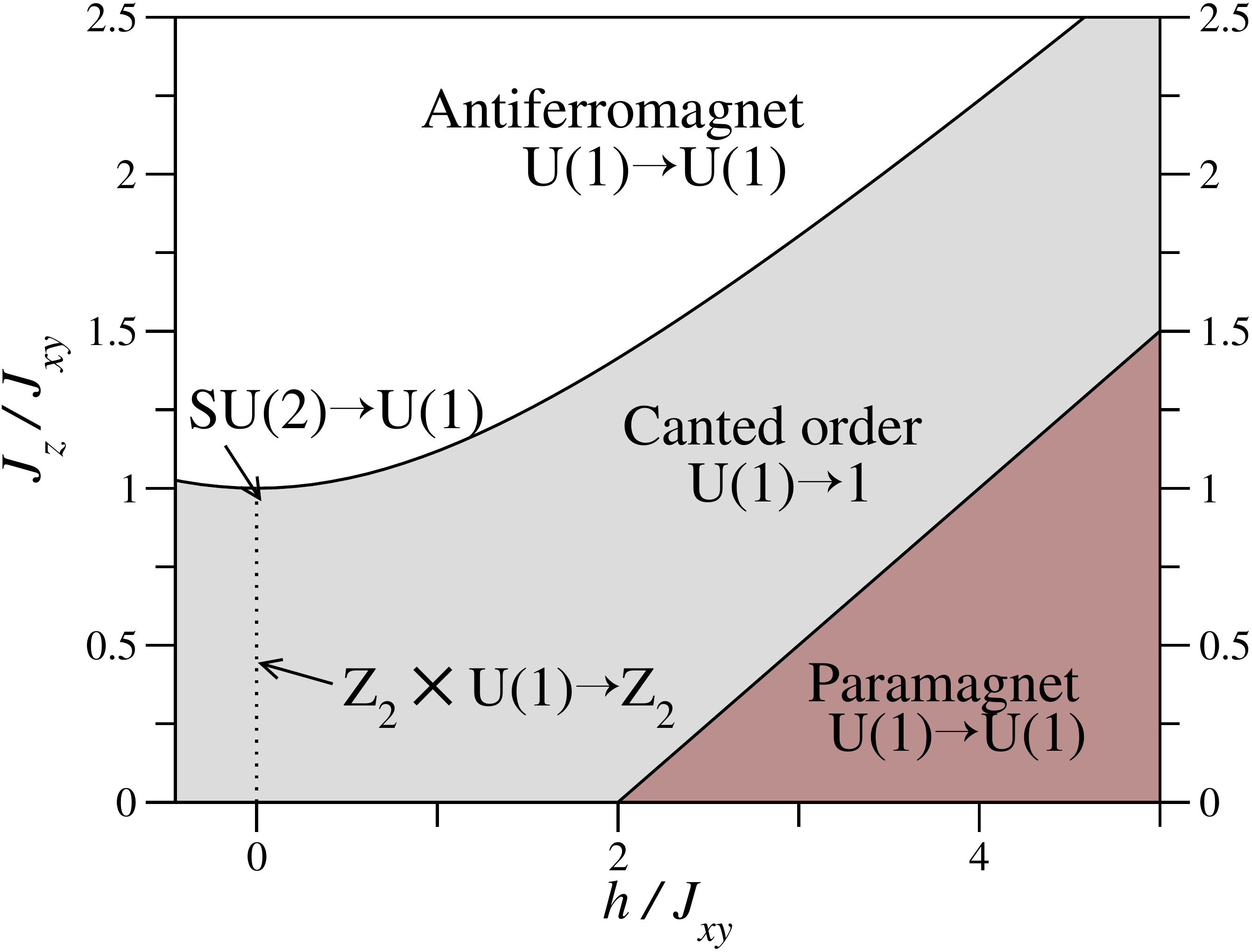} 
  \caption{Schematic phase diagram of the XXZ model for $J_{xy} = 1, h \geq 0, J_z \geq 0$. Indicated are the symmetries of the Hamiltonian and how they are broken in the respective phase. The phase diagram is symmetric with respect to the line $h=0, J_z \geq 0$. The transition from the antiferromagnet to the canted phase is of first order and simultaneously changes the $Z_2$ symmetry breaking from a staggered to a uniform pattern and breaks the $U(1)$ symmetry. The transition from the canted phase to the paramagnet is of second order and restores $U(1)$ symmetry. The dotted line from $h=J_z=0$ to $h=0, J_z=1$ does not indicate a phase transition and is meant as guide to the eye.\label{fig:pd_schematic}}
\end{figure}

The Hamiltonian of the XXZ model for spin-$\frac{1}{2}$ degrees of freedom is given by
\begin{equation} \label{eqn:XXZ}
H = \sum_{\langle i,j \rangle} \left( \frac{J_{xy}}{2}(\sigma_i^+ \sigma_j^- + \sigma_i^- \sigma_j^+) + J_z \sigma_i^z \sigma_j^z \right) + h \sum_i \sigma_i^z
\end{equation}
where $\sigma^x,\sigma^y,\sigma^z$ denote Pauli matrices and $\sigma^+ = \sigma^x + i \sigma^y$, $\sigma^- = \sigma^x - i \sigma^y$. By identifying $a_i = \frac{\sigma_i^+}{2}$, $a^{\dagger}_i = \frac{\sigma_i^-}{2}$, $n_i = \frac{\sigma_i^z+1}{2}$, we find (up to a constant) the Bose-Hubbard model in the limit of hard-core bosons, given by
\begin{equation}
H = -t \sum_{\langle i,j \rangle} (a_i^{\dagger} a_j + a_j^{\dagger} a_i) + V \sum_{\langle i,j \rangle} n_i n_j + \mu \sum_i n_i,
\end{equation}
where we use $J_{xy} = -\frac{t}{2}, J_z = \frac{V}{4}, h = \frac{\mu+V}{2}$.

\begin{figure}[t] 
   \centering
   \includegraphics[width=8cm]{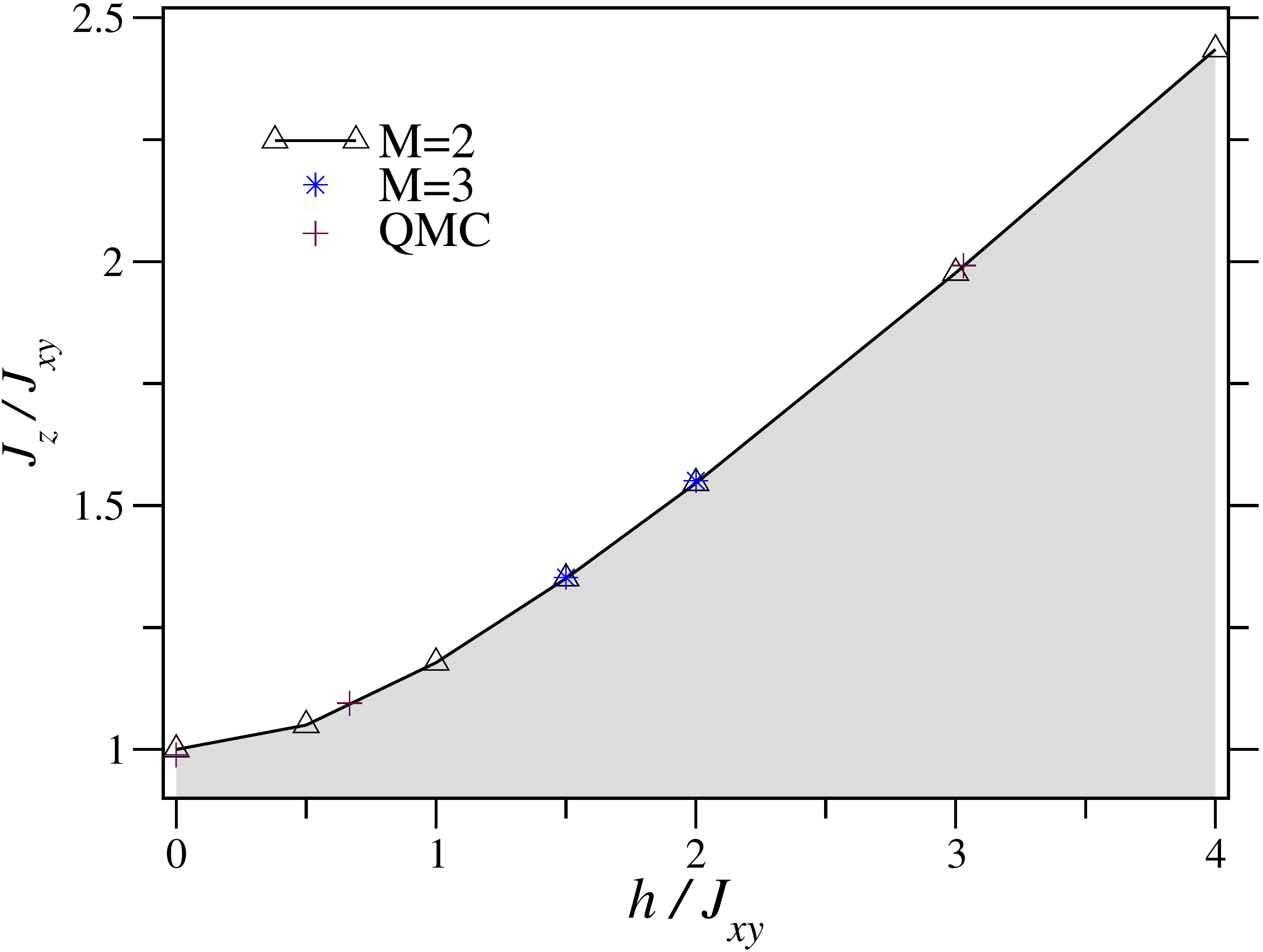} 
  \caption{Spin-flop transition of the XXZ model. Lines are guides to the eye and interpolate the results for an iPEPS with bond dimension 2. The QMC results are taken from Yunoki \cite{yunoki2002}.\label{fig:spinflop}}
\end{figure}

\begin{figure}[t]
  \centering
  \includegraphics[width=8cm]{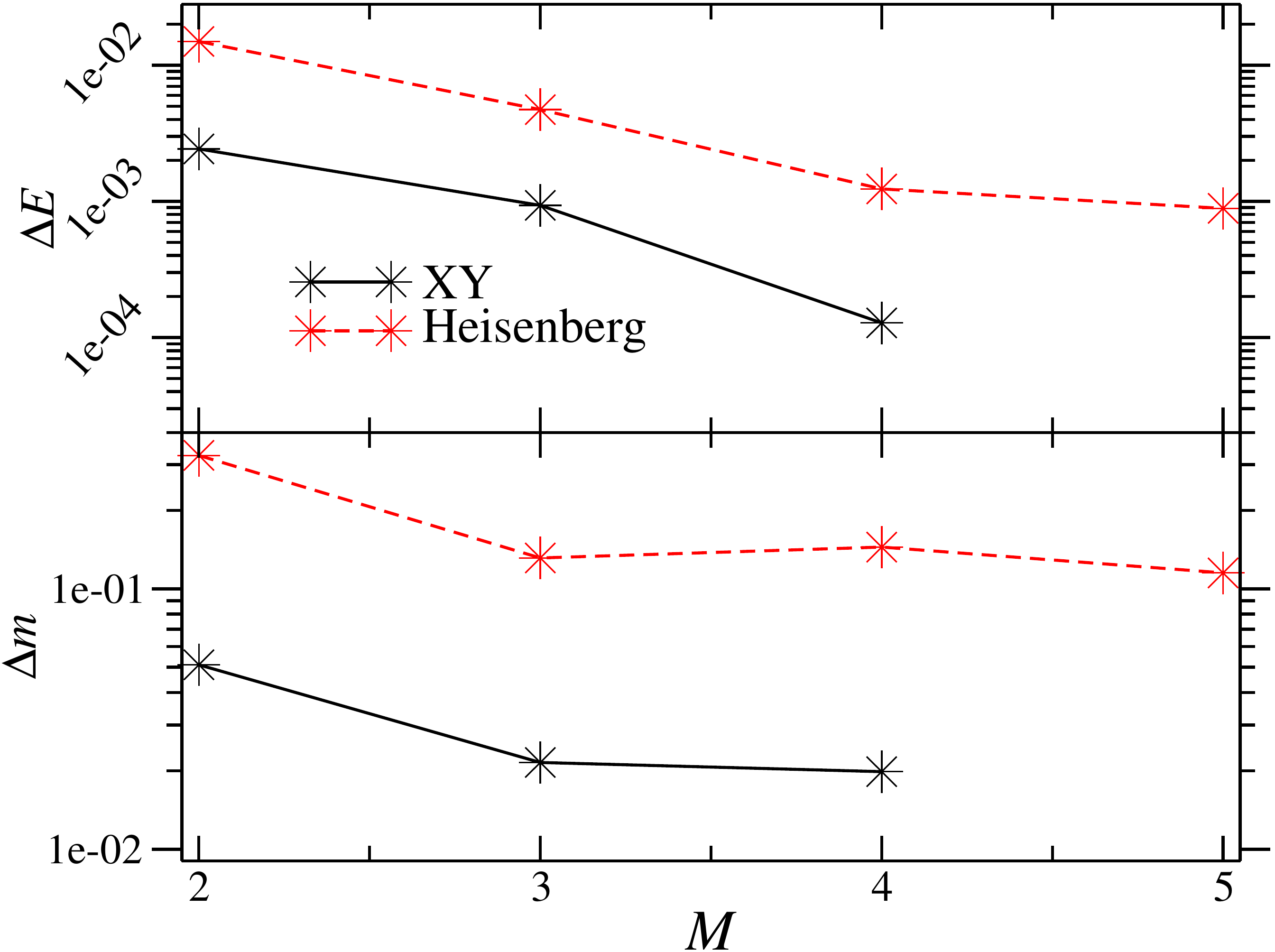}
  \caption{Convergence of relative errors $\Delta x = \vert x(\text{iPEPS}) - x(\text{QMC}) \vert / \vert x(\text{QMC}) \vert$ with bond dimension $M$. Monte Carlo results from \cite{sandvik1997,sandvik1999} are taken as exact. For the definition of the magnetization refer to the text.\label{fig:xy_xxx}}
\end{figure}

\begin{figure}[t]
  \centering
  \includegraphics[width=8cm]{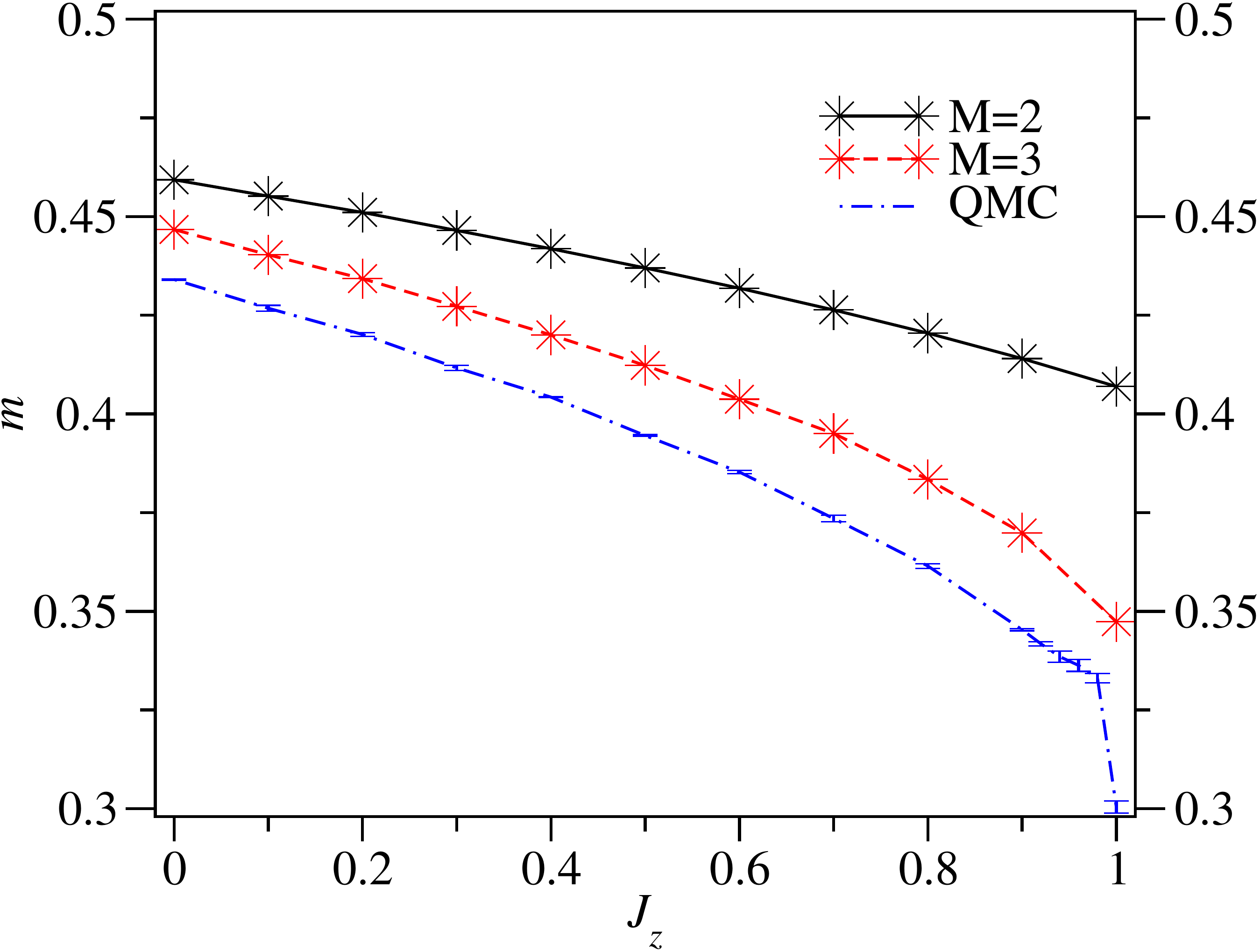}
  \caption{Magnetization $m$ (cf Eqn. (\ref{eqn:magnetization})). The error increases significantly as the isotropic Heisenberg model is approached. The restoration of the full $SU(2)$ symmetry at that point leads to a significant decrease of the magnetization, which is not correctly captured by iPEPS for the $M$ we considered. Finite size scaling was performed for the Monte Carlo calculations. \label{fig:ToHBP}}
\end{figure}

The phase diagram of this model has been studied to much detail and is well understood \cite{troyer2001,troyer2002,yunoki2002}. A sketch along with the symmetries of the Hamiltonian and the ground state in each phase is shown in Fig. \ref{fig:pd_schematic}. In the following discussion, XY and Z identify the axes in spin space, not in real space. Three different phases can be identified:
\begin{description}
\item[Paramagnet] A large magnetic field forces the system into a paramagnetic state, corresponding to a full or empty state in the bosonic picture. The $U(1)$ symmetry in the XY plane is not broken.
\item[Canted phase] This state, referred to as superfluid in the bosonic language, spontaneously breaks the $U(1)$ symmetry. At the line $h = 0$, the $\mathbb{Z}_2$ Ising symmetry is restored. At the isotropic point $J_{xy} = J_z$, this symmetry is enlarged to a $U(1)$ symmetry.
\item[Antiferromagnet] The antiferromagnetic phase (checkerboard solid) is characterized by a different symmetry breaking pattern of the $\mathbb{Z}_2$ symmetry: instead of the order induced by the external field, an antiferromagnetic pattern emerges. The $U(1)$ symmetry is not broken.
\end{description}
Only the superfluid phase is gapless.

The transition from the full or empty solid to the superfluid phase is a second order phase transition which can be located analytically as the appearance of a single boson in the system. The nature of the transition from the superfluid to the checkerboard-solid phase, referred to as spin-flop transition, was under debate for some time until Monte Carlo simulations clearly established it as first-order \cite{troyer2002}.

The Hamiltonian has a $U(1)$ symmetry which is spontaneously broken in the superfluid phase. At $J_z = J_{xy}$, the symmetry is enlarged to $SU(2)$, which is also broken with long-range order occurring at $T=0$. The enlarged symmetry group leads to larger quantum fluctuations, thereby reducing the staggered magnetization as the isotropic point is approached. As opposed to calculations on finite systems, the infinite system will fully break the symmetry, thereby allowing a direct calculation of the magnetizations from local observables.

The results we obtain for the spin-flop transition are shown in Fig. \ref{fig:spinflop}. The results for $M=2$ already match very well with the QMC results. The change with $M=3$ is very minor; at $h=1.5$, the relative difference between the critical coupling is on the order of $0.1 \%$. A very small $M$ is therefore sufficient to characterize this phase transition with very good accuracy.

\subsection{XXX and XY model} \label{sct:xxxxy}
A family of gapless systems is obtained for $J_{xy} = 1, h = 0, J_z \in [0,1]$. In Fig. \ref{fig:xy_xxx}, results for the energy and the spontaneous magnetization are shown, where the magnetization $m$ is defined as (subscript indices denote the site in the unit cell)
\begin{eqnarray} \label{eqn:magnetization}
m &=& \frac{1}{2} \langle m_k \rangle_{\text{unit cell}} \\
m_k &=& \sqrt{\sum_{i = x,y,z} \langle \sigma^i_k \rangle ^2}.
\end{eqnarray}
$\langle \sigma^z \rangle$ is vanishes for all $J_z < J_{xy}$. The data clearly shows that while energies are captured very well for both models, the large quantum fluctuations away from the N\'{e}el order, in particular for the isotropic case, are not captured by a PEPS of small dimension very well, leading to a significant error in the magnetization. Interestingly, the energy for the Heisenberg model decreases strictly monotonously with increasing $M$, while the magnetization does not improve significantly from $M=3$ to $M=4$ (this can also be seen in Fig. \ref{fig:dimer_op}). While the discrepancy between the improvement in energy and magnetization is clearly due to the vanishing gap above the ground state, the reason for the plateau in the magnetization cannot be determined conclusively.

Figure \ref{fig:ToHBP} shows how the decrease of the magnetization as $J_z$ is tuned from 0 to $J_{xy}$ is captured by the PEPS; in particular the rapid decrease as the $SU(2)$ symmetry is restored is not seen in the PEPS calculation.

\begin{figure}[t]
 \centering
 \includegraphics[width=8cm]{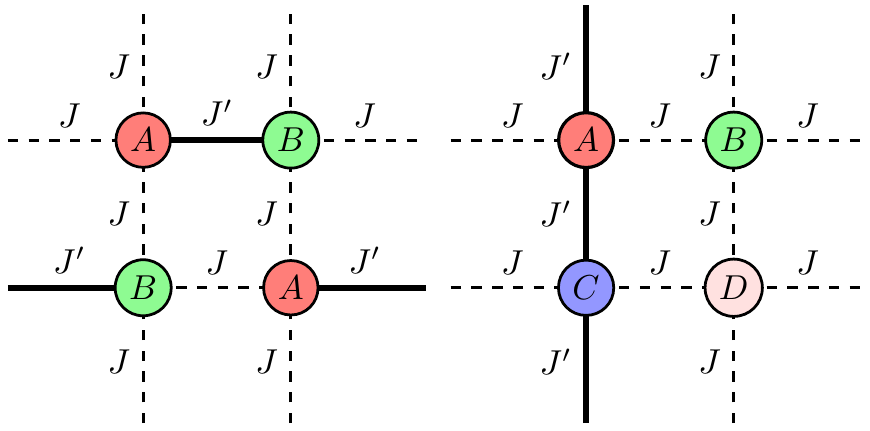}
 \caption{(color online) Different dimerization and unit cell patterns. The pattern on the left, which we refer to as \emph{staggered} pattern, is described by a unit cell with two independent tensors and used for the dimerized Heisenberg model. For $J'=0$, the honeycomb lattice is obtained. The second pattern, referred to as \emph{columnar} pattern, is used in the frustrated cases, but requires 4 instead of two independent tensors. The thick, solid bonds carry couplings $J_{xy}',J_z'$, the dashed lines carry couplings $J_{xy},J_z$. \label{fig:dimerization}}
\end{figure}

\subsection{Dimerized Heisenberg model} \label{sct:dimerized_hb}
\begin{figure}[t]
 \centering
 \includegraphics[width=8cm]{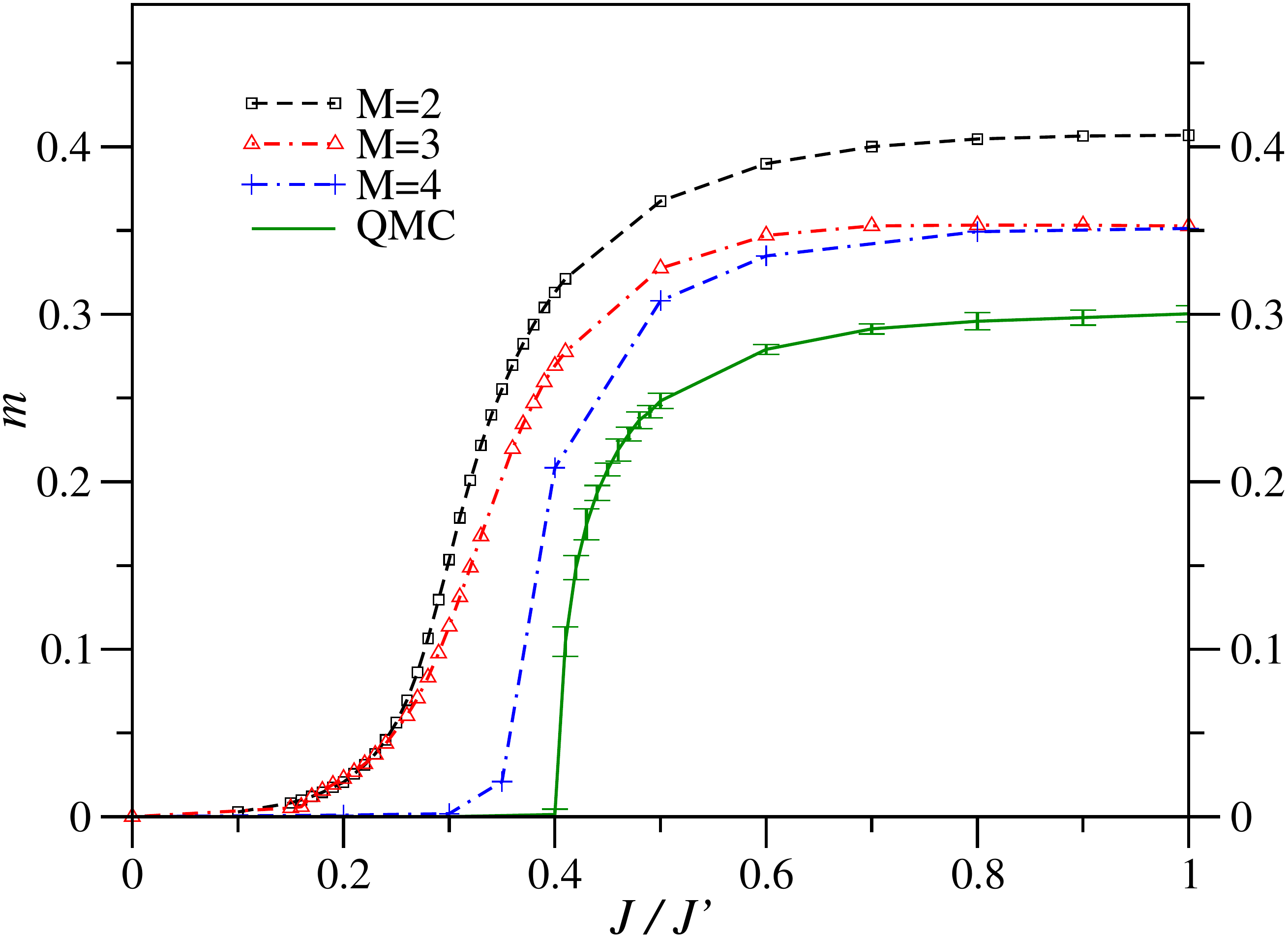}
 \caption{(color online) Staggered magnetization of the dimerized Heisenberg model as a function of the coupling ratio. $J/J' = 1$ corresponds to the isotropic Heisenberg model, while $J/J' = 0$ corresponds to isolated dimers on every second horizontal bond. \label{fig:dimer_op}}
\end{figure}

In order to test the accuracy of iPEPS for a quantum phase transition, we consider a dimerized Heisenberg model given by the Hamiltonian
\begin{equation}
H = \sum_{\langle i,j \rangle} J^{\langle i,j \rangle} \vec{\sigma}_i \vec{\sigma}_j,
\end{equation}
where we choose the couplings $J^{\langle i,j \rangle}$ inhomogenously according to the pattern shown on the left in Fig. \ref{fig:dimerization}. For $J'=1,J=0$, the state is made up of singlets on the strong-coupling bonds. In the limit $J=J'$, the isotropic Heisenberg model is recovered. Between those limits, a second-order phase transition occurs \cite{wenzel2008}. The behaviour of the order parameter, the staggered magnetization, is shown in Fig. \ref{fig:dimer_op}. While the results do indicate a second-order phase transition, the critical point observed for small $M$ deviates significantly from the Monte Carlo result.

This is a clear indication that the phase on both sides of the transition is a highly entangled one: even in the limit of pure dimers, $M=2$ is required to describe the ground state; in the limit of the Heisenberg models, quantum fluctuations around the N\'{e}el state are very strong. It is interesting to note that the results for $M=3$ interpolates between those for $M=2$ towards the dimerized phase and $M=4$ towards the antiferromagnet.

\section{Frustrated Heisenberg Models} \label{sct:frust}

%\emph{Alternative motivation: It is well-known that thermal fluctuations destroy long-range order for low-dimensional quantum systems. For some cases, it is also known that quantum fluctuations destroy the long-range order and give rise to disordered spin liquid phases. Another mechanism that can lead to the destruction of long-range order is the formation of local singlets as exemplified by the dimerized Heisenberg model discussed above.}

%\emph{This model easily allows the introduction of frustration to study both effects simultaneously. By choosing the signs of the couplings $J, J'$ in Fig. \ref{fig:dimerization} such that $J J' < 0$, we obtain a family of frustrated systems that can be studied with iPEPS in a straightforward manner. We first perform mean-field calculations to obtain an initial understanding of the phase diagram; then, quantum fluctuations are introduced by foobaring doodoo. [...]}

To study the applicability of the iPEPS method to frustrated quantum models, we have studied a phase transition in the family of models obtained by choosing the signs of the different couplings as shown in Fig. \ref{fig:dimerization} in such a way that the product around each plaquette is negative, i.e. $J J' < 0$. The class of models allows a large degree of freedom in the choice of parameters. In particular, we can choose full Heisenberg coupling $J_{xy}^{\langle i,j \rangle} = J_z^{\langle i,j \rangle}$ or restrict the coupling to the XY plane, $|J_{xy}| > 0, J_z = 0$.

The case of Heisenberg couplings on the staggered pattern has been studied by Kr\"{u}ger et al \cite{krueger2000} using the Coupled-Cluster Method and Exact Diagonalization. The authors find a phase transition from the antiferromagnetic state in the non-frustrated limit to a hexatic phase. Classically, this occurs at a coupling ratio $J'/J = 1/3$; in the quantum model, the antiferromagnetic phase is stable up to $J'/J \approx 1.35$. While the phase transition is second-order in the classical case, it is conjectured to be of first order in the quantum case. The hexatic phase for this pattern however has a large magnetic unit cell which renders it unsuitable for simulation with iPEPS. In the following discussion, we will therefore focus on the case of XY coupling on the columnar configuration with $J = -1$.

Classical analogues of spin-$\frac{1}{2}$ systems can be obtained by replacing the classical spins with Ising ($\mathbb{Z}_2$), XY ($O(2)$) or Heisenberg ($O(3)$) spins. Equivalently, one can consider mean-field solutions of the quantum system since for spin-$\frac{1}{2}$ degrees of freedom, the mean-field approximation to a Hamiltonian of the form
\begin{equation}
H = \sum_{\langle i,j \rangle} J^{\langle i,j \rangle}_x \sigma_i^x \sigma_j^x + J^{\langle i,j \rangle}_y \sigma_i^y \sigma_j^y + J^{\langle i,j \rangle}_z \sigma_i^z \sigma_j^z,
\end{equation}
can be mapped exactly to a classical Hamiltonian for Heisenberg spins. We can therefore choose to either solve a classical system or perform a mean-field simulation, e.g. by minimizing the energy of an iPEPS state with bond dimension $M=1$. In this case, the renormalization procedure described above reduces to the multiplication with a scalar which cancels for all physical observables.

The classical system has been studied by Villain \cite{villain1977} and solved for the case of $|J'| = |J|$ with Ising and XY spins. For XY spins on the staggered pattern, a two-fold degenerate hexatic state is found, which exhibits a $2 \times 2$ site unit cell. Using numerical mean-field calculations on a $2 \times 2$ unit cell, we locate a second-order phase transition from a ferromagnetically ordered state to the hexatic state at $J'/J = 1/3$. Our result for $J'/J = 1$ complies with the result by Villain.

\begin{figure}[t]
 \centering
 \includegraphics[width=8cm]{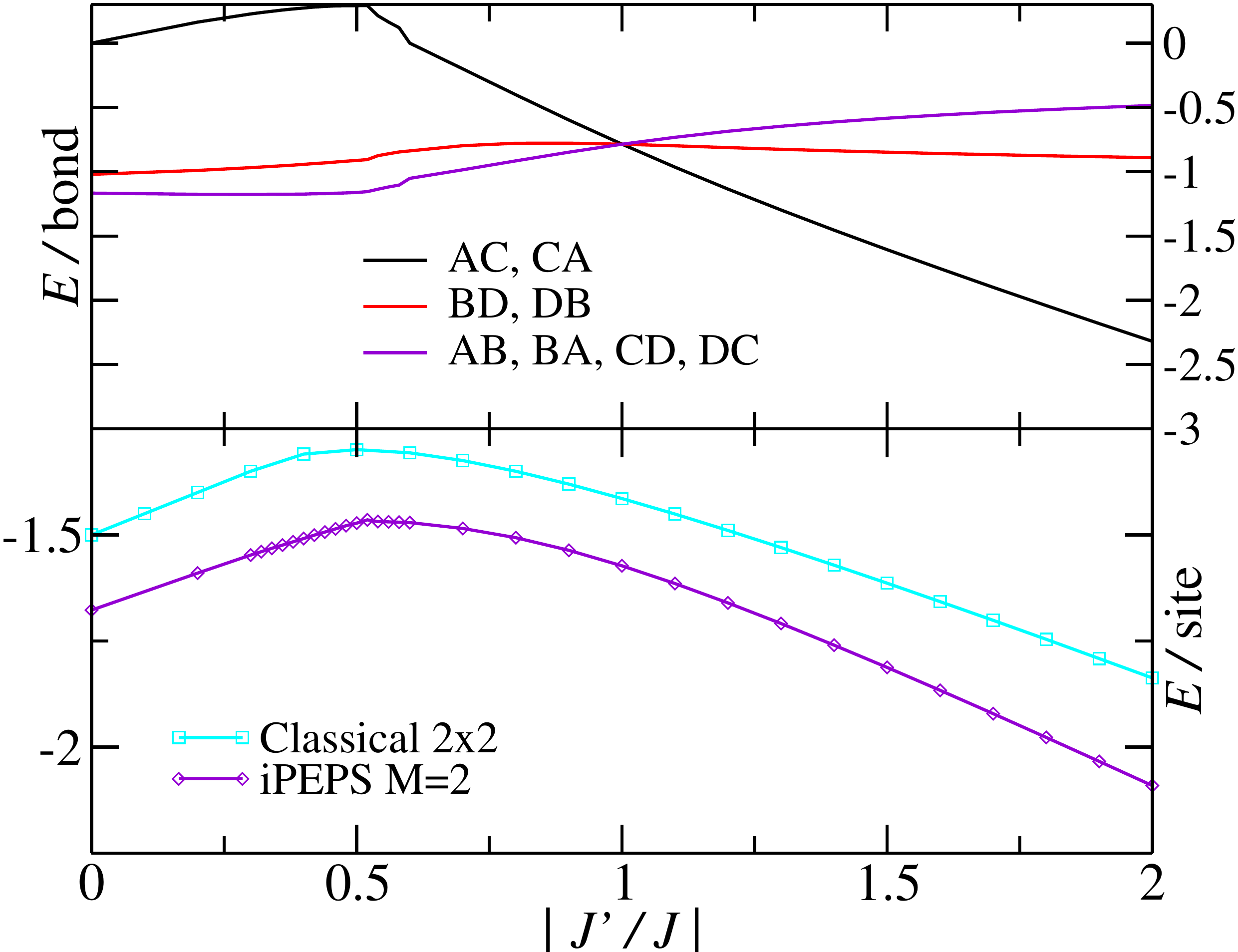}
 \caption{(color online) Energies for the different bonds and comparison of energies for the classical and quantum columnar frustrated XY model. \label{fig:energies}}
\end{figure}

\begin{figure}[t]
 \centering
 \includegraphics[width=8cm]{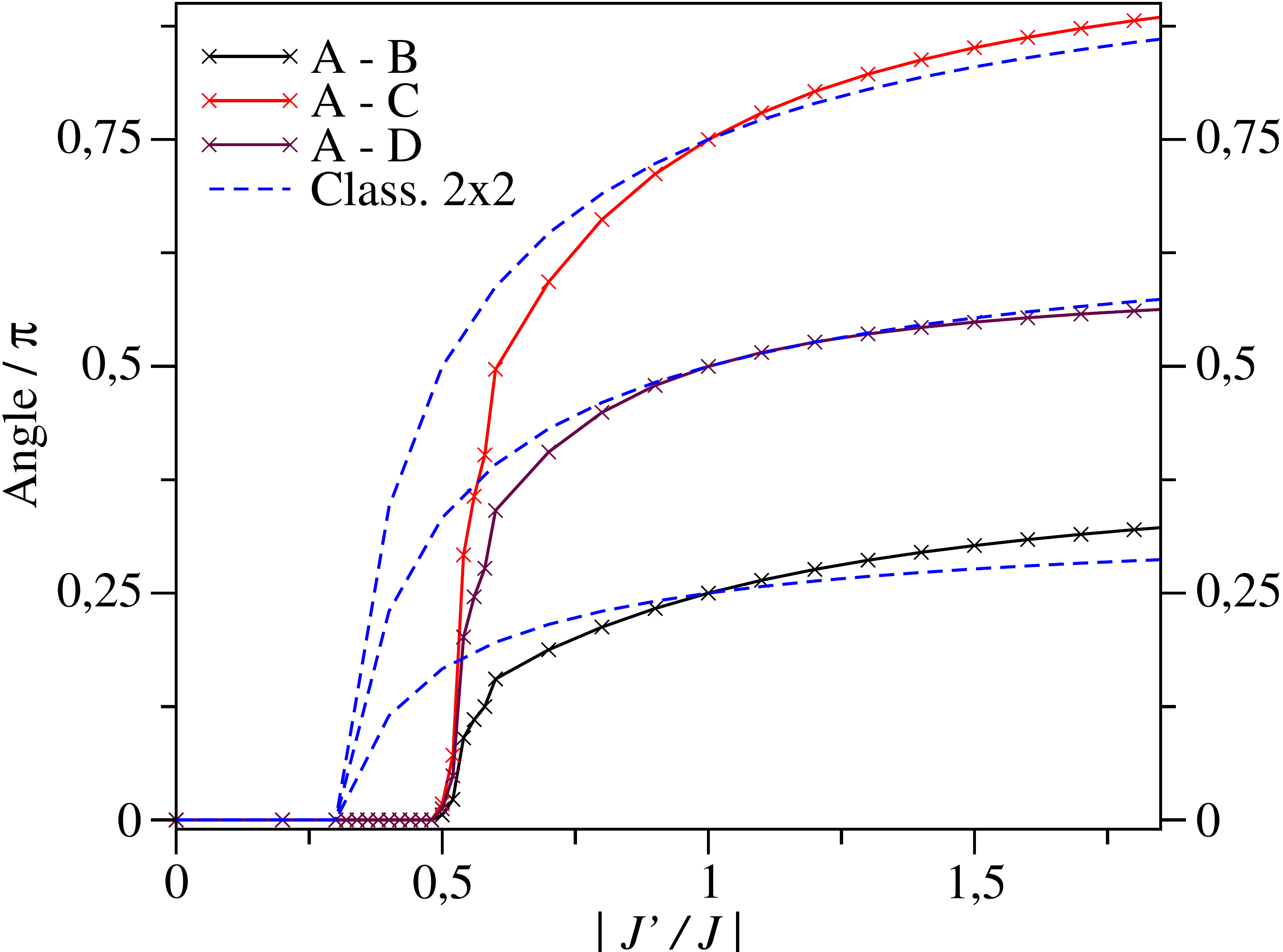}
 \caption{(color online) Relative angles between the magnetizations in the XY plane for the quantum and classical columnar frustrated XY model, which is equivalent to a mean-field solution of the quantum model. Notice that at maximal frustration, the angles for the classical and quantum case coincide. The transition is shifted from $J'/J = 1/3$ to a larger value of $J'/J \approx 0.45$. \label{fig:angles}}
\end{figure}

\begin{figure}[t]
 \centering
 \includegraphics[width=1.8in]{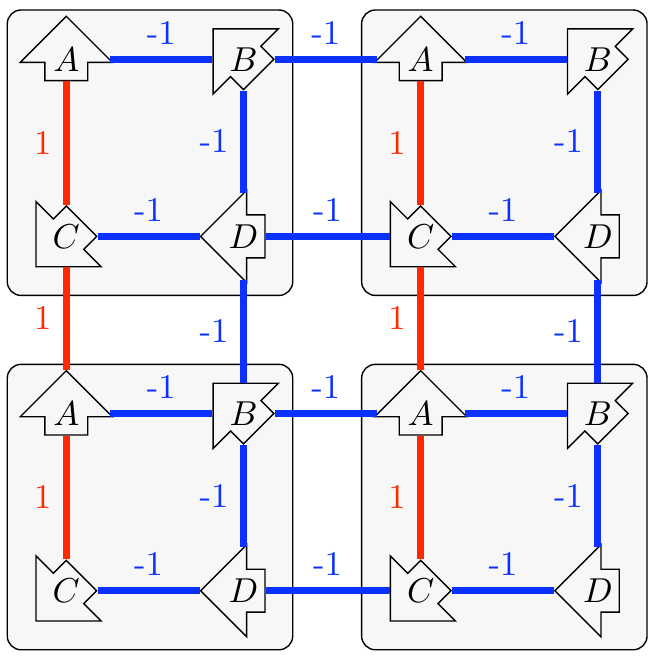}
 \caption{(color online) Ground state of the columnar-pattern frustrated model with $J_{xy} = -1$ and $J'_{xy} = 1$. Four magnetic unit cells are shown for illustration purposes. The arrows indicate the magnetization in the XY plane (up to global rotation). The angles of the quantum ground state are identical to those of the classical ground state. \label{fig:order}}
\end{figure}

To simulate the quantum system beyond mean-field, we use an iPEPS ansatz with $M=2$ and a $2 \times 2$ unit cell. The classical ground state is used as initial state for the iPEPS simulation, since this improves stability and convergence speed of the algorithm. We measure the energy on each bond and the relative angles between the spins in the XY plane. For the ferromagnetic configuration, these are $0$; for $\angle(AC) = \angle(AB) = \pi, \angle(AD) = 0$, the state is antiferromagnetic. In between these limits, a hexatic state is found. Therefore, the angles can be interpreted as order parameters for the phase transition.

In Fig. \ref{fig:energies}, the energy for each bond and a comparison of the total energy per site to the classical result is shown. For the non-frustrated case, $J' = 0$, we can also compare to Quantum Monte Carlo results. The energy found with iPEPS for $M=2$ is $E_0 = -1.676$, while the Monte Carlo results for small lattices extrapolate to $E_0 = -1.683(1)$. The relative error of $\Delta E = 0.0042$ is comparable to that obtained for the XY model on the square lattice.

We find that the magnetization always remains in the XY plane and is reduced by quantum fluctuations to a value on the order of $m \approx 0.45$; at all points except the maximally frustrated point $J' = J$, the magnetization is different on the AD and BC sublattices. The angles as a function of the coupling ratio are shown in Fig. \ref{fig:angles}. They indicate a phase transition analogous to the classical one, which however is shifted towards a higher critical coupling of around $J'/J = 0.45$. This agrees with previous results indicating that quantum fluctuations stabilize the collinear order in this class of models.

The state at the point of maximal frustration, $|J'/J|=1$, is shown in Fig. \ref{fig:order}. The state corresponds directly to Villain's solution up to a reduction of the magnetization on each site to $m \approx 0.451$. We have cross-checked this result by starting the simulation with different initial states.

In order to investigate the order of the phase transition, we can study the stability of a state upon quenches of the parameters that take the system across the phase transition as discussed in Sect. \ref{sct:fo}. In our simulations, we do not observe such stability of the state, indicating that the transition remains continuous in the presence of quantum fluctuations.

\section{Conclusion}
We have confirmed that the iPEPS method can be applied with excellent accuracy to first-order phase transitions using the example of the spin-flop transition in the hard-core boson model. The stability of a homogeneous system across the phase transition allows us to clearly distinguish the nature of the phase transition and determine the location of the transition very accurately.

For the isotropic Heisenberg model and a dimerized Heisenberg model, the accuracy compared to finite-size extrapolations of unbiased Monte Carlo simulations decreases significantly. This can be understood as a signature of large quantum fluctuations away from the product state, which cannot be captured by a PEPS of small bond dimension.

For a simple frustrated model on the square lattice, we have shown that the method converges to results that agree with expectations based on mean-field simulations and previous results obtained using the Coupled-Cluster Method. Whereas the applicability of path integral Monte Carlo methods to such systems is limited by the so-called \emph{sign problem}, the accuracy of iPEPS is only limited by the entanglement of the ground state and frustration does not introduce additional difficulties. We therefore expect that the method will become a valuable tool for such systems.

During completion of this work, we became aware of another study of the spin-flop transition using an algorithm based on tensor-product states by P. Chen et al \cite{chen2009}.

%\section*{Acknowledgements}
\ack
We would like to thank F. Verstraete, S. Wenzel, S. Trebst and J.I. Cirac for useful discussions. Calculations were performed on the ETH Brutus cluster.

\section*{References}
%\bibliography{bibliography}{}
%\bibliographystyle{unsrt}

\end{document}